\newif\ifprp
 \prptrue   

\ifprp
 \documentstyle[twoside,aps,prb,epsf,floats]{revtex} 
\else
 \documentstyle[aps,prb]{revtex}
\fi

\begin{document}

\ifprp
 \twocolumn[\hsize\textwidth\columnwidth\hsize\csname 
 @twocolumnfalse\endcsname 
\fi

\draft

\title{Weak  Localization  in  Semiconductor Multi-Quantum Well Structures}

\author{F. G. Pikus}

\address{State University of New York, Stony Brook, NY 11794-3800}

\author{G. E. Pikus}

\address{A.F. Ioffe Physicotechnical  Institute 194021 St  Petersburg,
RUSSIA}

\date{\today} 
\maketitle 

\begin{abstract}
We have studied the phenomenon of weak localization in multi-quantum-well (MQW)
structures in the regime of weak tunneling, when superlattice minibands 
are not formed. We have calculated the effect of weak localization on
conductivity, which in this situation is described by a system of coupled
Dyson equations. The tunneling across the MQW structure is found, in general, 
to suppress the weak localization effect on conductivity. 
\end{abstract}

\ifprp
 \vskip 2pc ] 
\fi


\narrowtext

The weak localization, which  results  from  interference of two
electron waves propagating along a closed path  in
opposite directions and manifests itself, among other effects, in the 
phenomenon of the negative magnetoresistance, 
has been studied extensively in metals and
semiconductors, as well as thin films and quantum wells of these
materials\cite{WLReviews,Altshuler80,Hikami80,Altshuler81}. 
Recently, a lot of attention has been devoted
to studies of the weak localization in more complex systems, such as
superlattices, multiple quantum wells, quantum wires, 
etc.\cite{WLSuperlatTheory,WLSuperlatExp,WLWeirdSystems}

The physics of the weak localization in the superlattices is of
particular interest, and has been fairly extensively investigated, both
theoretically\cite{WLSuperlatTheory} and experimentally\cite{WLSuperlatExp}. 
The theories vary in their approach, however, they all start from
(implicit or explicit) assumption that the electron spectrum consists of
well-defined minibands. This supposition is in fact equivalent to 
assuming that the tunneling time between two wells $\tau_{12}$ is much
shorter than the momentum lifetime for in-plane motion, $\tau_{11}$.
This is a necessary condition for miniband formation, and for 
the structure to be treated as a superlattice.

The opposite case, when $\tau_{11} \ll \tau_{12}$ and 
an electron diffuses in an single quantum well for a long time
before tunneling into a neighboring well, has never been studied, to the
best of our knowledge. However, in practice both cases can be easily
implemented, the first one - in a superlattice, the second one - in
weakly coupled multiple quantum wells. 
In this latter case, there are no
minibands, and no coherent motion across the planes of the MQW structure. 
The tunneling is sequential, i.e. there is no tunneling through many
barriers at once, and tunneling into the next well does not depend on
the history of electron motion. 

We will be interested in the effects of such weak tunneling on the weak
localization in lateral transport. Consequently, we choose the structure
and make assumptions which help us to investigate this effect without
additional complications. For example, we take the simple model for
electron spin relaxation, which is described by the single
spin-relaxation time $\tau_{\rm SO}=\tau_{s_{xx}} = \tau_{s_{yy}}
=2\tau_{s_{zz}}$\cite{LinearTerms}. From 
Refs.~[\onlinecite{Knap96,Hassenkam97}] one can see that in $\rm A_3B_5$
quantum wells this approximation works quite well for large electron
densities. We also assume
that the temperature is low, $T \ll \epsilon_{\rm F}$, and that
$\hbar/\tau_{11}\epsilon_{\rm F} \ll 1$ - the usual conditions in the theory
of weak localization\cite{Altshuler80}i (here $\epsilon_{\rm F}$ is the
electron Fermi energy).

The weak localization contribution to the conductivity of a single 
quantum well is given by the
well-known expression\cite{Hikami80,Altshuler81}:

\begin{equation}
\Delta \sigma_1 = - {e^2 D  \over \pi \hbar} \cdot
2 \pi \nu_0 \tau_{11}^2 
\sum_{\alpha \beta}
\int\limits_0^{q_{\max}} {\rm  \kern.24em \vrule
width.05em
height1.4ex depth-.05ex \kern-.26em C}_{\alpha \beta \beta \alpha}({\bf q}) \,
\frac{d^2 q}{(2 \pi)^2},
\end{equation}

\noindent where
 
${\rm \kern.24em \vrule  width.05em
height1.4ex depth-.05ex \kern-.26em C}(\bf q)_{\alpha
\beta \beta \alpha}$ is
the Cooperon, $\alpha$ and $\beta$ are the spin indices which we will
omit in all subsequent expressions,
$q_{\max}^2 = (D t_1)^{-1}$, $D =  v^2 t_1/2$ is 
the diffusion coefficient,  $t_1$ is the transport time for in-plane
motion in a well (differs from the lifetime $\tau_{11}$ for long-range
scattering), and
$\nu_0 =  m/2\pi$ is  the density  of states  at the  Fermi level. 

\ifprp
  \begin{figure}[t]
  \epsfxsize=3.5in
  \epsffile{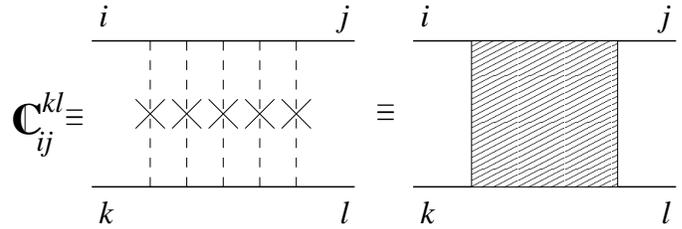}
  \caption{Cooperon for multi-quantum-well structure, with indices $i$, $j$, $k$,
  and $l$ denoting quantum well numbers.}
  \label{Fig:Cooperon}
  \end{figure}
\fi

In the case of a MQW structure, the Cooperon is a 4-dimensional matrix
indexed by the well numbers, ${\rm \kern.24em \vrule  width.05em
height1.4ex depth-.05ex \kern-.26em C}_{ij}^{kl}$ (see
Fig.~\ref{Fig:Cooperon}). The total conductivity of the structure with $N$
wells is 

\begin{eqnarray}
\Delta \sigma &=& \sum_{n=1}^N \Delta \sigma_n 
\label{SigmaSum}
\\
&=& - {e^2 D  \over \pi \hbar}
\cdot
2 \pi \nu_0 \tau_{11}^2 
\sum_{\alpha \beta}
\int\limits_0^{q_{\max}} 
\sum_{n=1}^N {\rm  \kern.24em \vrule width.05em
height1.4ex depth-.05ex  \kern-.26em C}_{nn} ({\bf q}) \,
\frac{d^2 q}{(2 \pi)^2}, 
\nonumber
\end{eqnarray}

\noindent where ${\rm  \kern.24em \vrule width.05em
height1.4ex depth-.05ex  \kern-.26em C}_{nn} \equiv 
{\rm  \kern.24em \vrule width.05em
height1.4ex depth-.05ex  \kern-.26em C}_{nn}^{nn}$. 
For large number of wells $N$ the calculations can be greatly simplified
if we impose periodical boundary conditions, so the $N-$th well is
connected with the 1st. We will also assume all wells 
to be identical. In this case

\ifprp
  \begin{figure*}[t]
  \epsfxsize=6.5in
  \epsffile{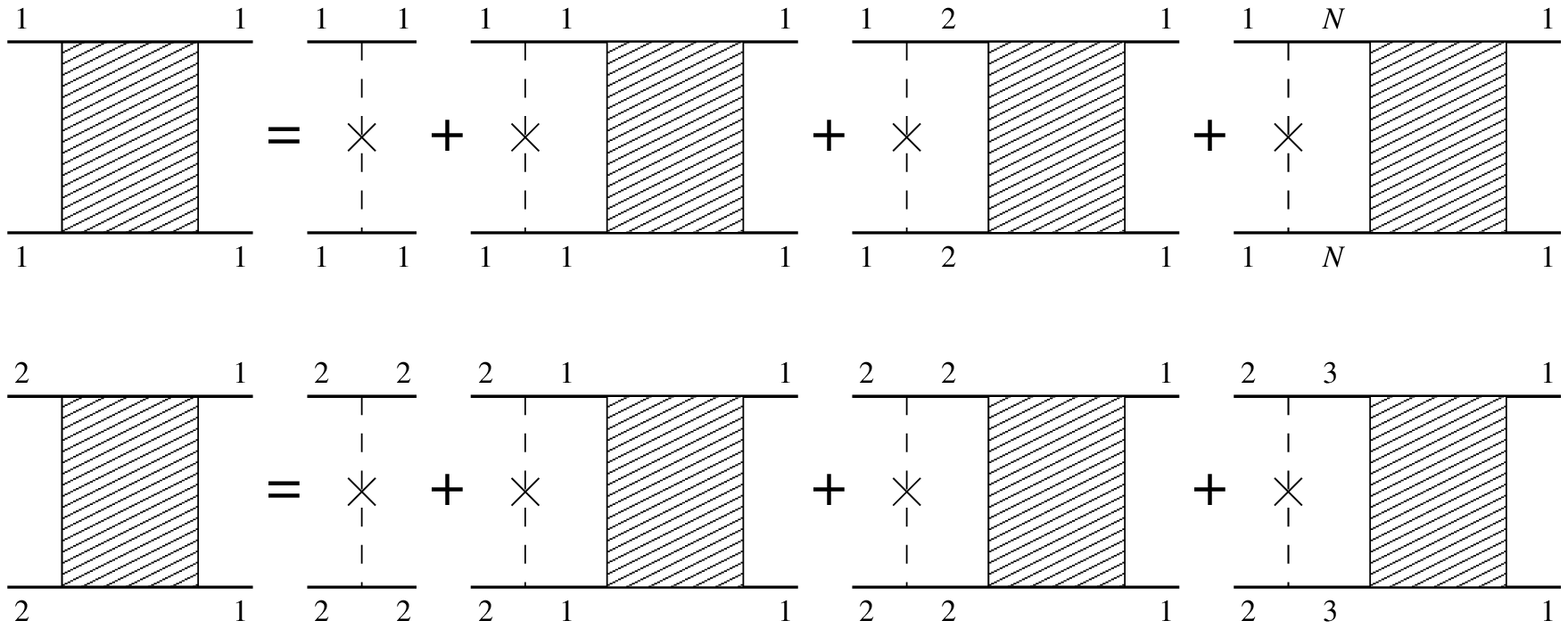}
  \caption{The system of Dyson equations for
  ${\rm  \kern.24em \vrule width.05em
  height1.4ex depth-.05ex  \kern-.26em C}_{11}$,
  ${\rm  \kern.24em \vrule width.05em
  height1.4ex depth-.05ex  \kern-.26em C}_{21}$, etc.
  in diagrammatic representation. Only first two equations are shown.}
  \label{Fig:Dyson}
  \end{figure*}
\fi

\begin{eqnarray}
\Delta \sigma &=& N \Delta \sigma_1 
\label{SigmaN} \\
&=& - {e^2 D  \over \pi \hbar}
\cdot
2 \pi \nu_0 \tau_{11}^2 
\sum_{\alpha \beta}
\int\limits_0^{q_{\max}}
N {\rm  \kern.24em \vrule width.05em
height1.4ex depth-.05ex  \kern-.26em C}_{11} ({\bf q}) \,
\frac{d^2 q}{(2 \pi)^2}, 
\nonumber
\end{eqnarray}

The Cooperon can be found from the Dyson equation, or, in our case,
system of Dyson equations, since the component 
${\rm  \kern.24em \vrule width.05em
height1.4ex depth-.05ex  \kern-.26em C}_{nm}$
is connected with ${\rm  \kern.24em \vrule width.05em
height1.4ex depth-.05ex  \kern-.26em C}_{n m\pm1}$. These equations are
represented by diagrams shown in Fig.~\ref{Fig:Dyson}. Note that each
``component'' is actually a 4-dimensional $2\times2\times2\times2$ 
matrix with spin indices, which can alternatively be represented as a
$4\times4$ matrix indexed by pairs $\alpha \beta$.
These
equations contain only components ${\rm  \kern.24em \vrule width.05em
height1.4ex depth-.05ex  \kern-.26em C}_{1n} \equiv 
{\rm  \kern.24em \vrule width.05em
height1.4ex depth-.05ex  \kern-.26em C}_{1n}^{1n}$;
similar equations exist for components 
${\rm  \kern.24em \vrule width.05em
height1.4ex depth-.05ex  \kern-.26em C}_{2n}$ and so on.
There are, of
course, other components of the Cooperon ${\rm  \kern.24em \vrule
width.05em
height1.4ex depth-.05ex  \kern-.26em C}_{ij}^{kl}$,
however, none of them is relevant for the weak
localization conductivity correction: some turn to 0 after 
averaging over impurities (like 
${\rm  \kern.24em \vrule width.05em
height1.4ex depth-.05ex  \kern-.26em C}_{11}^{21}$
which contains $\langle V_{11} V_{21} \rangle = 0$,
where $V_{ij}$ is the matrix element for scattering with initial state
in well $i$ and final state in well $j$); 
other, while non-zero, do not contribute to to the conductivity, 
(for example, the contribution of ${\rm  \kern.24em \vrule width.05em
height1.4ex depth-.05ex  \kern-.26em C}_{11}^{22}$ is proportional to
the overlap of the wave functions in different wells, because lines 1
and 2 would have to meet in the vortex, and we assume that this overlap
is extremely small). These ignored diagram also does not contribute in the Dyson
equations, unless the scattering in the two wells is correlated, i.e.
they do not couple with ${\rm  \kern.24em \vrule width.05em
height1.4ex depth-.05ex  \kern-.26em C}_{11}$ and 
${\rm  \kern.24em \vrule width.05em
height1.4ex depth-.05ex  \kern-.26em C}_{12}$ as long as $\langle V_{11}
V_{22} \rangle = 0$. Another component, 
${\rm  \kern.24em \vrule width.05em
height1.4ex depth-.05ex  \kern-.26em C}_{12}^{21}$, in principle can
give contribution to the conductivity (just like its spin analog, ${\rm
\kern.24em \vrule width.05em
height1.4ex depth-.05ex  \kern-.26em C}_{\alpha \beta}^{\beta \alpha}$,
does give such contribution). However, in order for this component to be
non-zero, there has to be at least one scattering act which occurs on
the same scattering center but in different wells. We assume that the
scattering centers have short enough range so there is no correlation in
the scattering potentials in neighboring wells. Hence, there exists no
such center which can scatter electrons in both wells, and the Cooperon
component ${\rm  \kern.24em \vrule width.05em
height1.4ex depth-.05ex  \kern-.26em C}_{12}^{21}$ vanishes.

Written as a system of integral equations, Fig.~\ref{Fig:Dyson} becomes

\begin{eqnarray}
&&{\rm  \kern.24em \vrule width.05em
height1.4ex depth-.05ex  \kern-.26em C}_{11}({\bf q}) =
\left|V_{11}\right|^2 
\nonumber \\
&+& \int d^2 g_1 \, \left|V_{11}\right|^2 G^+_1(\omega, {\bf g}_1 + {\bf q})
                                 G^-_1(\omega, -{\bf g}_1)
{\rm  \kern.24em \vrule width.05em
height1.4ex depth-.05ex  \kern-.26em C}_{11}({\bf q}) 
\nonumber \\
&+& \int d^2 g_2 \, \left|V_{12}\right|^2 G^+_2(\omega, {\bf g}_2 + {\bf q})
                                 G^-_2(\omega, -{\bf g}_2)
{\rm  \kern.24em \vrule width.05em
height1.4ex depth-.05ex  \kern-.26em C}_{21}({\bf q})
\nonumber \\
&+& \int d^2 g_N \, \left|V_{1N}\right|^2 G^+_N(\omega, {\bf g}_N + {\bf q})
                                 G^-_N(\omega, -{\bf g}_N)
{\rm  \kern.24em \vrule width.05em
height1.4ex depth-.05ex  \kern-.26em C}_{N1}({\bf q}),
\nonumber \\
&&{\rm  \kern.24em \vrule width.05em
height1.4ex depth-.05ex  \kern-.26em C}_{21}({\bf q}) =
\left|V_{21}\right|^2 
\label{Dyson} \\
&+& \int d^2 g_1 \, \left|V_{21}\right|^2 G^+_1(\omega, {\bf g}_1 + {\bf q})
                                 G^-_1(\omega, -{\bf g}_1)
{\rm  \kern.24em \vrule width.05em
height1.4ex depth-.05ex  \kern-.26em C}_{11}({\bf q}) 
\nonumber \\
&+& \int d^2 g_2 \, \left|V_{22}\right|^2 G^+_2(\omega, {\bf g}_2 + {\bf q})
                                 G^-_2(\omega, -{\bf g}_2)
{\rm  \kern.24em \vrule width.05em
height1.4ex depth-.05ex  \kern-.26em C}_{21}({\bf q})
\nonumber \\
&+& \int d^2 g_3 \, \left|V_{23}\right|^2 G^+_3(\omega, {\bf g}_3 + {\bf q})
                                 G^-_3(\omega, -{\bf g}_3)
{\rm  \kern.24em \vrule width.05em
height1.4ex depth-.05ex  \kern-.26em C}_{31}({\bf q}),
\nonumber
\end{eqnarray}

\noindent and so on for all the components ${\rm  \kern.24em \vrule width.05em
height1.4ex depth-.05ex  \kern-.26em C}_{n1}$. Note that the tunneling
is only possible between the neighboring wells, so only terms with
$V_{n,n\pm1}$ are present.
Here $G^{\pm}(\omega, {\bf k})$  are  the Green's
functions,

\begin{equation}
G^{\pm}(\omega, {\bf k}) = \displaystyle
\frac{1}{\omega - E(k) 
\pm i\left(\frac{1}{2\tau_{11}} + \frac{1}{2\tau_\varphi}\right)},
\end{equation}

\noindent $E(k) = k^2/2m$ and $\tau_\varphi$ is the phase
relaxation time. After  the
integration by $E(g)$ in the right-hand side of Eq.~(\ref{Dyson})  the
result  is  expanded  up  to  second  order  terms  in series in small
parameters $\tau_{11}/\tau_\varphi$ and ${\bf vq} \tau_{11}$ (${\bf v}  =
\partial E/\partial {\bf k}$). Also, the Cooperon 
${\rm  \kern.24em \vrule width.05em
height1.4ex depth-.05ex  \kern-.26em C}_{ij}^{kl}({\bf q})$ 
is expanded in harmonics:

\begin{equation}
{\rm  \kern.24em \vrule width.05em
height1.4ex depth-.05ex  \kern-.26em C}_{ij}^{kl}({\bf q}) = 
\sum_n 
C_{ij}^{{kl}^{(n)}} \cos n \phi_{\bf q},
\end{equation}

\noindent where $\tan \phi_{\bf q} = q_y/q_x$. These transformations
follow closely Refs.~[\onlinecite{Altshuler80,Hikami80,Altshuler81}] and
are described in details, for example, in Ref.~[\onlinecite{Knap96}]. 
Only the 0-th harmonic of the Cooperon, $C_{ij}^{kl} \equiv 
C_{ij}^{{kl}^{(0)}}$, gives non-negligible contribution to the
conductivity. For the components $C_{n1}$  we arrive to the following
system of linear equations:

\begin{eqnarray}
C_{11} &=& \left|V_{11}\right|^2 + \frac{\tau_{0}}{\tau_{11}}\left(1 -
L_{11}\tau_{0}\right)C_{11} 
\nonumber \\
&+& \frac{\tau_{0}}{\tau_{12}}\left(1 -
L_{12}\tau_{0}\right)\left(C_{12} + C_{N1}\right), 
\nonumber \\
C_{21} &=& \left|V_{21}\right|^2 + \frac{\tau_{0}}{\tau_{11}}\left(1 -
L_{11}\tau_{0}\right)C_{21} 
\nonumber \\
&+& \frac{\tau_{0}}{\tau_{12}}\left(1 -
L_{12}\tau_{0}\right)\left(C_{11} + C_{31}\right), 
\label{System} \\
C_{31} &=& \frac{\tau_{0}}{\tau_{11}}\left(1 -
L_{11}\tau_{0}\right)C_{31} 
\nonumber \\
&+& \frac{\tau_{0}}{\tau_{12}}\left(1 -
L_{12}\tau_{0}\right)\left(C_{21} + C_{41}\right), ...
\nonumber
\end{eqnarray}

\noindent Here 

$$\frac{1}{\tau_0} = \frac{1}{\tau_{11}} + \frac{2}{\tau_{12}},$$

\noindent $\tau_{11}$ is the momentum lifetime in a well, 
$\tau_{12}$ is a tunneling time between two neighboring wells:

$$\tau_{1n}^{-1} = 2\pi\nu_0|V_{1n}|^2,$$

\noindent and 

\begin{equation}
L_{11}=L_{12} = L = D q^2 +\frac{1}{\tau_\phi} +
\frac{1}{\tau_{\rm SO}^m} 
\end{equation}

\noindent is a $4\times4$ matrix,indexed, like 
the Cooperon itself, by pairs of spin indices $\alpha \beta$. 
In the basis of eigenfunctions of total momentum of the pair of electrons,
singlet $\phi_0$ and triplet $\phi_1^m$ with $m = -1, 0, 1$ this matrix
becomes

\begin{equation}
L_0 = D q^2 + \frac{1}{\tau_\phi}, \
L_1^m = D q^2 + \frac{1}{\tau_\phi} + \frac{1+\delta_{m0}}{\tau_{\rm SO}},
\label{Lq}
\end{equation}

\noindent where $\delta_{m0}$ is the $\delta$-symbol.

The system (\ref{System}) can be further simplified if we exploit the 
condition that $\tau_{12} \gg \tau_{11} \approx \tau_{11}$, and keep only
terms of the first order in $\tau_{11}/\tau_{12}$ (however, the product 
$L\tau_{12}$ can be large or small, and we make no assumptions about its
magnitude here). We can also simplify the expressions by introducing

\begin{equation}
S(q) = 2\pi\nu_0\tau_{11}^2 C(q)
\end{equation}

\noindent Then the system of linear equations for $S_{1n}$ becomes

\begin{eqnarray}
A_1 S_{11} + A_2 (S_{21} + S_{N1}) &=& \tau_{11}, 
\nonumber \\
A_1 S_{21} + A_2 (S_{11} + S_{31}) &=& \tau_{12} \approx 0, 
\label{SystemS} \\
A_1 S_{31} + A_2 (S_{21} + S_{41}) &=& 0, \ ...
\nonumber 
\end{eqnarray}

\noindent where 

\begin{eqnarray}
A_1 &=& \frac{\tau_0}{\tau_{12}}\left(2 + L \tau_{12}\right) \approx
\frac{\tau_0}{\tau_{12}}(2 + x), 
\nonumber \\
A_2 &=& \frac{\tau_0}{\tau_{12}}\left(-1 + L \tau_0\right) \approx
-\frac{\tau_0}{\tau_{12}}, 
\label{A} \\
x &=& L \tau_{12}.
\nonumber
\end{eqnarray}

\noindent The system of linear equations (\ref{SystemS}) allows us to
find $S_{11}(q)$ (which is a $4\times4$ matrix in the basis of singlet
and triplet eigenfunctions $\phi_0$, $\phi_1^m$). 
The weak localization correction to the
conductivity (\ref{SigmaN}) can be expressed
through the eigenvalues of the inverse matrix, $S_{11}^{-1}(q)$, as 
follows\cite{Iordanskii94,Pikus95,Knap96}:

\begin{equation}
\Delta \sigma = - {e^2 D  \over \pi \hbar}
N \int\limits_0^{q_{\max}}
\left(-\frac{1}{E_0} + \sum_{m=-1}^1\frac{1}{E_1^m}\right)  
\frac{d^2 q}{(2 \pi)^2}.
\end{equation}

\noindent Here eigenvalue $E_0$ corresponds to the eigenfunction
$\phi_0$, and $E_1^m$ - to $\phi_1^m$. Since the matrix $L$, and, hence,
$S_{11}$, are diagonal in this basis, the eigenvalues $E_0$ and $E_1^m$
of $S_{11}^{-1}(q)$
are just inverse of its corresponding diagonal elements,
$(S_{11})_0$ and $(S_{11})_1^m$, respectively, and 

\begin{equation}
\Delta \sigma = - {e^2 D  \over \pi \hbar}
N \int\limits_0^{q_{\max}}
\left(-S_0 + \sum_{m=-1}^1S_1^m\right)
\frac{d^2 q}{(2 \pi)^2}.
\label{SigmaS}
\end{equation}

\noindent Here we have omitted indices $11$ of $S$ to simplify the
notation.

In a magnetic field $B$ perpendicular to the planes of MQW 
the wave vector $\bf q$ becomes an
operator with the commutation relations

\begin{equation}
[q_+q_-] = {\delta \over D}, \ \delta = {4 e B D \over \hbar c},
\label{Commut}
\end{equation}

\noindent where $q_\pm = q_x \pm i q_y$.
This  allows  us  to  introduce  creation  and annihilation
operators $a^\dagger$ and $a$, respectively, for which $[aa^\dagger] =
1$:

\begin{equation}
D^{1/2} q_+ = \delta^{1/2} a,
\quad
D^{1/2} q_- = \delta^{1/2} a^\dagger,
\quad
D q^2 = \delta \{a a^\dagger\}.
\label{OperQ}
\end{equation}

\noindent The non-zero matrix elements of these operators are

\begin{eqnarray}
\left\langle n-1 \right| a \left| n \right\rangle & = &
\left\langle n \right| a^\dagger \left| n-1 \right\rangle = \sqrt{n},
\nonumber \\
\left\langle n \right| \{a a^\dagger\} \left| n \right\rangle
& = & n + {1 \over 2}.
\end{eqnarray}

\noindent The integration over $q$ in a magnetic field becomes summation
over $n$, and the weak localization correction to  the conductivity can
be written as

\begin{equation}
\Delta \sigma = - {e^2 \delta \over 4 \pi^2 \hbar} N \sum_{n=0}^{n_{\max}}
\left(-S_{0n} + \sum_{m=-1}^1S_{1n}^m\right).
\label{NMRB}
\end{equation}

\noindent where $n_{max} = 1/\delta\tau_{11}$. It is convenient to
extend the summation in Eq.~(\ref{NMRB}) to $n \rightarrow \infty$. One
can do it using the following transformation, which exploits the
condition $n_{\max} \gg 1$ (this is the necessary condition for the
diffusion approximation to work in the first place). We will see below
that at large $n$ the asymptotic values of $S_{0n}$ and $S_{1n}^m$ are
$1/\delta n$. Therefore, the expression under the sum in Eq.~(\ref{NMRB}) falls
off as $2/\delta n$ and the sum diverges at large $n$. To remove this
divergence we add and subtract $2/n+1$ to each term in Eq.~(\ref{NMRB}):
$\delta \sum_{n=0}^{n_{\max}} \left(-S_{0n} + \sum_{m=-1}^1S_{1n}^m\right) = 
\sum_{n=0}^{n_{\max}} \left(-\delta S_{0n} + \delta \sum_{m=-1}^1S_{1n}^m
- \frac{2}{n+1}\right) + \sum_{n=0}^{n_{\max}}\frac{2}{n+1}$.
The first sum can be extended to $n=\infty$ because the
expression under the sum falls of as $1/n^2$ at large $n$. The second
sum, again for $n_{\max} \gg 1$ can be approximated by $\ln n_{\max}$. Since
the quantity of practical interest is not $\Delta\sigma$ itself but the
magnetoconductivity $\Delta\sigma(B) - \Delta\sigma(0)$, which is
measured experimentally, we can replace $\ln n_{\max} = \ln
1/\delta\tau_{11}$ by $\ln 1/\delta\tau_\varphi$ and ignore the
$B$-independent term $\ln \tau_\varphi/\tau_{11}$. After these
transformations, we arrive to the expression for the magnetoconductivity
which does not contain $\tau_{11}$ at all:

\begin{eqnarray}
\Delta\sigma(B) &-& \Delta\sigma(0) = 
\label{NMRBinf} \\
&-& {e^2 \over 4 \pi^2 \hbar} 
N \left[\sum_{n=0}^\infty \left(-\delta S_{0n} + \delta \sum_{m=-1}^1S_{1n}^m -
\frac{2}{n+1}\right)\right.
\nonumber \\
&-&\left. 2 \ln(\delta\tau_\varphi)\right].
\nonumber
\end{eqnarray}

The linear system (\ref{SystemS}) with coefficients Eq.~(\ref{A})
remains the same in a magnetic field, with $L$ now being a function of
$n$ instead of $q$:

\begin{eqnarray}
L_{0n} &=& \delta \left(n + \frac{1}{2}\right) + \frac{1}{\tau_\phi}, \
\label{Ln} \\
L_{1n}^m &=& \delta \left(n + \frac{1}{2}\right) + 
\frac{1}{\tau_\phi} + \frac{1+\delta_{m0}}{\tau_{\rm SO}},
\nonumber
\end{eqnarray}

The solution of the system (\ref{SystemS}) can be written in the
following form, for each $n$ and each spin index $l, m$:

\begin{equation}
S \equiv S_{11} = \frac{1}{L} F_N(x),
\label{SLF}
\end{equation}

\noindent where $N$ is the number of wells in the MQW structure and $x =
L\tau_{12}$. The function $F_N(x)$ can be found by solving the system
(\ref{SystemS}) for any particular $N$. One can check by direct
substitution that the following general expression is valid for any $N$:

\begin{eqnarray}
F_N(x) &=& \frac{1}{2}(-x)^{N/2} \, 
{}_2F_1\left(\frac{1-N}{2}, 1-\frac{N}{2}; 1-N; -\frac{4}{x}\right)
\nonumber \\
&\times&
\sec\left(N \arccos\frac{\sqrt{-x}}{2}\right),
\end{eqnarray}

\noindent where ${}_2F_1(a, b; c; z)$ is the Hypergeometric function.
For practical purposes it is more convenient to represent $F_N(x)$ by a
ratio of two polynomials:

\begin{eqnarray}
F_N(x) &= (x+4)
\frac{\displaystyle
\sum_{m=0}^{N/2} x^{N/2-m} \frac{N}{N-m} {m \choose N-m}
}{\displaystyle
\sum_{m=0}^{N/2-1} x^{N/2-m-1} {m \choose N-m-1}
}, \ &{\rm even}\ N,
\nonumber \\
\label{FN} \\
F_N(x) &=
\frac{\displaystyle
\sum_{m=0}^{(N-1)/2} x^{(N-1)/2-m} {m \choose N-m-1}
}{\displaystyle
\sum_{m=0}^{(N-1)/2} x^{(N-1)/2-m} {m \choose N-m}
}, \ &{\rm odd}\ N,
\nonumber
\end{eqnarray}

\noindent It is interesting to note that this expression has a very
compact limit at $N\rightarrow\infty$. By expanding $F_N(1/y)$ for very
large $N$ in series around $y=0$ one can see that the first $N$ coefficients
of the series do not change with increasing $N$. They are given by the
following expression:

\begin{equation}
F_N\left(\frac{1}{y}\right) = \sum_{i=0}^{N-1} K_i y^i + ..., \qquad
K_i = \frac{(-4)^i \Gamma\left(i + \frac{1}{2}\right)}
{\sqrt{\pi} n!}.
\label{FSeries}
\end{equation}

\ifprp
  \begin{figure}[t]
  \epsfxsize=3.5in
  \epsffile{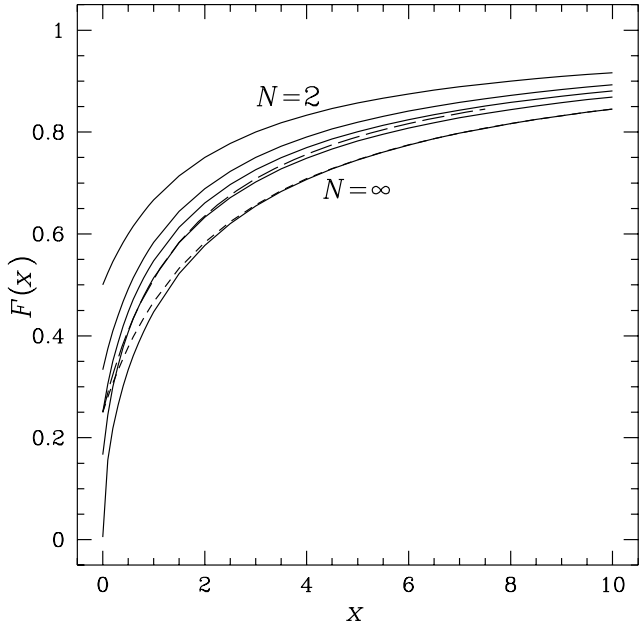}
  \caption{Function $F^*_N(x)$, which determines Cooperon components entering the 
  expressions for conductivity {\bf (}see Eq.~(\ref{SLF}){\bf )} for $N=2$, 3, 
  4, 6, and $N=\infty$ (solid curves, from highest to lowest). All functions 
  shown by solid curves are obtained without using periodical boundary conditions.
  Using these boundary conditions for $N=4$ results in $F_4(x)$
  shown by short-dash line. Note 
  that the periodical boundary conditions become much more accurate if the $x$
  axis is scaled by $(N-1)/N$ (for the case of $N=4$, shown by long-dashes line,
  the abscissae axis was scaled by $3/4$.}
  \label{Fig:F}
  \end{figure}
\fi

\noindent To obtain the limit $F_\infty(x)$ we extend the summation to 
infinity, and the sum can be written in a closed form:

\begin{equation}
F_\infty(x) = \frac{1}{\sqrt{\displaystyle 1 + \frac{4}{x}}}
\label{FInf}
\end{equation}

In the above derivation of $F_N(x)$ we have assumed the periodical boundary 
conditions, i.e. that electron can tunnel between the 1st and $N$th wells. 
While greatly simplifies the calculations, this condition becomes inaccurate
for small $N$, and a more accurate solution is needed. If there is no 
periodical boundary conditions, each well gives a different contribution 
into the total conductivity {\bf (}see Eq.~\ref{SigmaSum}){\bf)}. The
components of the cooperon $C_{nn}$ can be found from $N$ systems of linear
equations, system number $m$ is written for coefficients $C_{1m}$, 
$C_{2m}$, ... , $C_{Nm}$, and is similar to Eq.~\ref{System} 
but does not contain the tunneling between 1st and $N$th wells. 
From these systems we can find the components of the Cooperon $C_{nn}$, or 
$S_{nn}$, the sum of which enters the expression for conductivity. We can
define an ``average'' $S^*_{11}$ to be
used instead of $S_{11}$ in the above expressions for conductivity, 
and, similarly to Eq.~(\ref{SLF}), introduce function $F^*_N(x)$:

\begin{equation}
S^* = 1/N \cdot \sum_{n=1}^N S_{nn} = \frac{1}{L} F^*_N(x).
\end{equation}

Figure~\ref{Fig:F} shows the function $F^*_N(x)$ for 
different number of wells $N$, from 2 to infinity. For large $N$,
the difference between $F^*_N(x)$ and $F_N(x)$ is negligible,
and the two coincide exactly for $N = \infty$. However, for small $N$, 
the periodical boundary conditions 
introduce a noticeable error; for example, in Fig.~\ref{Fig:F} 
we compare the functions $F^*_4(x)$ and $F_4(x)$ (short-dashed curve).
It is interesting to note that replacing $\tau_{12}$ by an ``effective'' 
tunneling time $\tau_{12}^\prime = N/(N-1) \cdot \tau_{12}$ 
(or $x$ by $x^\prime = N/(N-1) \cdot x$)
makes the function $F_N(x^\prime)$ almost coincide with $F^*_N(x)$. 
The reason for this behaviors is intuitively clear: by applying periodic boundary conditions we have added an ``extra'' tunneling link to the structure, in
addition to the $N-1$ ``real'' links. Scaling $\tau_{12}$ makes the ``cumulative''
tunneling rate for the structure with periodical boundary conditions equal 
to that of the real structure. For 
$N=2$ this procedure, obviously, gives exactly correct function, since
the periodic boundary conditions at $N=2$ are equivalent to just doubling the 
tunneling rate. In Fig.~\ref{Fig:F} we show by the long-dashed curve 
the scaled function $F_4(4/3x)$; one can see that it looks very similar to 
$F^*_4(x)$.

We now have all the necessary expressions, namely, Eqs.~(\ref{NMRBinf},
\ref{Ln}, \ref{SLF}, \ref{FN}), to calculate the
magnetoconductivity due to the weak localization in an MQW structure. 
First, following [\onlinecite{Iordanskii94,Pikus95}] we introduce the 
characteristical magnetic fields\cite{tau12}

\begin{eqnarray}
H_\varphi &=& {c  \hbar \over 4 e D  \tau_\varphi}{\rm ,} \ \
\frac{B}{H_\varphi} = \delta\tau_\varphi, 
\nonumber \\
H_{\rm SO} &=& {c \hbar \over 4 e D \tau_{\rm SO}}{\rm ,} \ \
\frac{H_\varphi}{H_{\rm SO}} = \frac{\tau_{\rm SO}}{\tau_\varphi}.
\label{HChar} \\
H_{12} &=& {c \hbar \over 4 e D \tau_{12}}{\rm ,} \ \
\frac{H_\varphi}{H_{12}} = \frac{\tau_{12}}{\tau_\varphi}.
\nonumber
\end{eqnarray}

\noindent and dimensionless conductivity {\em per one quantum well}:

\begin{equation}
\delta\sigma(B) = \frac{2 \pi \hbar}{Ne^2}
\left[\Delta\sigma(B) - \Delta\sigma(0)\right].
\label{SigmaDim}
\end{equation}

We begin by analyzing two limiting cases of weak and strong tunneling,
or $x \gg 1$ and $x \ll 1$, respectively. In the first case we can
immediately see from Eq.~(\ref{FSeries}) that $F_N(\infty) = 1$, so $S =
1/L$ (note that at large $n$ we also have $x \gg 1$ and, therefore, 
$S = 1/L \sim 1/\delta n$). 
Substituting expressions for $L$ from Eq.~(\ref{Ln}) we obtain
the standard result for single quantum well\cite{Altshuler80}:

\begin{eqnarray}
\delta\sigma(B) &=& {1 \over 2 \pi}
\left[
-\Psi\left(\frac{1}{2} + \frac{H_\varphi}{B}\right)
+2\Psi\left(\frac{1}{2} + \frac{H_\varphi}{B} +
  \frac{H_{\rm SO}}{B}\right)\right.
\nonumber \\
&+&\left.\Psi\left(\frac{1}{2} + \frac{H_\varphi}{B} +
  \frac{2H_{\rm SO}}{B}\right)
-2\ln\left(\frac{H_\varphi}{B}\right),
\right]
\end{eqnarray}

\noindent where $\Psi$ is the digamma-function.
This is not surprising since $x \rightarrow
\infty$ corresponds to $\tau_{12} \rightarrow \infty$ in which case the
parallel wells become completely independent. In the opposite case, $x
\rightarrow 0$, we have from Eq.~(\ref{FN}) that $F_N(0)=1/N$, so $N S = 1/L$
and we again obtain the classical result, only now for the entire
MQW structure since the tunneling couples the wells so that they all act
as one well.

\ifprp
  \begin{figure}[t]
  \epsfxsize=3.5in
  \epsffile{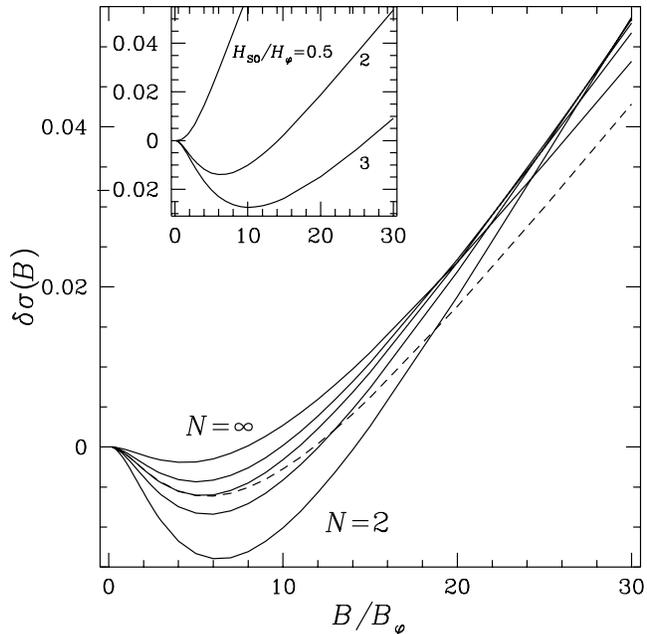}
  \caption{Dimensionless magneconductivity $\delta\sigma(B)$ for 
  $H_{\rm SO}/H_\varphi = 2$ and $H_{12}/H_\varphi = 4$ and different number 
  of wells $N=\infty$, 6, 4, 3, and $N=2$ (solid lines, in order from higher 
  to lower near the minimum). The solid lines were calculated without using 
  the periodical boundary conditions. The magnetoconductivity calculated with 
  these conditions is shown for $N=4$ by a dashed line.
  The inset shows the effect of spin relaxation on magnetoconductivity for $N=2$
  and $H_{12}/H_\varphi = 4$.}
  \label{Fig:ManyN}
  \end{figure}
\fi

We now present the results of our calculations of the weak localization 
correction to the conductivity in multiple quantum wells for various 
parameters: number of wells $N$, spin relaxation time $\tau_{\rm SO}$, 
and tunneling time $\tau_{12}$. We begin by analyzing the dependence 
of the dimensional magnetoconductivity $\delta\sigma(B)$ on the 
number of wells $N$, which is shown in Fig.~\ref{Fig:ManyN} for 
$H_{\rm SO}/H_\varphi = 2$ (this value is typical for $\rm A_3B_5$ quantum
wells, see Ref.~[\onlinecite{Knap96}]), 
$H_{12}/H_\varphi = 4$, and $N$ from 2 to infinity.
This figure also shows the effect of periodical boundary conditions at 
small $N$: the magnetoconductivity calculated with such boundary conditions 
for $N=4$ is shown by a dashed curve.
One can see that increasing the number of wells makes the minimum on the
magnetoconductivity curve less pronounced. 
It is known\cite{Iordanskii94,Pikus95} that increasing $\tau_{\rm SO}$ 
has similar effect on the magnetoconductivity, the minimum becomes less 
pronounced and eventually disappears as $\tau_{\rm SO}$ increases; we 
illustrate this effect in the insert of Fig.~\ref{Fig:ManyN} where
$\delta\sigma(B)$ is shown for $N=2$, $H_{12}/H_\varphi = 4$, and 
$H_{\rm SO}/H_\varphi = 1/2$ (top curve), $H_{\rm SO}/H_\varphi = 2$ 
(middle curve), and $H_{\rm SO}/H_\varphi = 3$ (the lowest curve). 
One can say that coupling the wells tighter effectively makes the spin
relaxation slower. 

\ifprp
  \begin{figure}[t]
  \epsfxsize=3.5in
  \epsffile{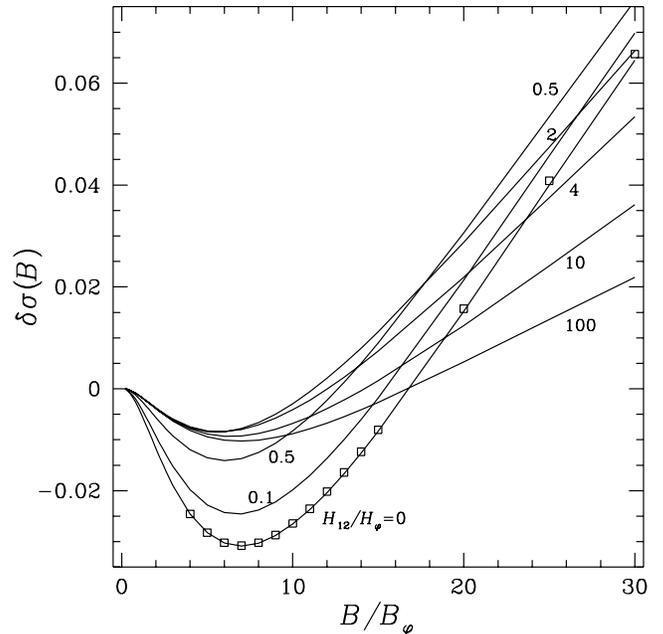}
  \caption{Dimensionless magnetoconductivity $\delta\sigma$ for $N=3$,
  $H_{\rm SO}/H_\varphi=2$, and $H_{12}/H_\varphi$ from 0 to 100. The 
  squares show $3\cdot \delta\sigma(B, H_{12}/H_\varphi=100)$ (which is
  very close to the limit at $H_{12} = \infty$), to illustrate
  the connection Eq.~(\ref{Sigma0Inf}) between magnetoconductivity at
  $H_{12}=0$ and $H_{12} = \infty$.}
  \label{Fig:N3}
  \end{figure}
\fi

This similarity between effects of increasing number of wells and increasing 
$\tau_{\rm SO}$ is not complete: one can see from Fig.~\ref{Fig:ManyN} 
that the minimum on the magnetoconductivity curve does not disappear as $N$ 
increases, while increasing $\tau_{\rm SO}$ leads to a monotonic 
$\delta\sigma(B)$.
After studying the dependencies of $\delta\sigma(B)$ on all the parameters,
we have found that
if for a given $\tau_{\rm SO}$ and some $N$, $\tau_{12}$ 
the function $\delta\sigma(B)$ has a minimum, this will not change for
any other $N$ and $\tau_{12}$ as long as $\tau_{\rm SO}$ is maintained 
the same. To better understand this behavior, we examine the 
magnetoconductivity dependence on $\tau_{12}$. Figure~\ref{Fig:N3} 
shows the magnetoconductivity for $N = 3$, 
$H_{\rm SO}/H_\varphi = 2$, and $H_{12}/H_\varphi$ from 0 to 100.
From the analysis of the limiting cases $\tau_{12} \ll \tau_\varphi$ and
$\tau_{12} \gg \tau_\varphi$ we can conclude that 

\begin{equation}
\delta\sigma(B, \tau_{12}=\infty) = 
N \delta\sigma(B, \tau_{12}=0).
\label{Sigma0Inf}
\end{equation}

\noindent It is obvious that both limiting cases are either monotonic, or not.
From Fig.~\ref{Fig:N3} one can see that, while it is not true that all 
$\delta\sigma(B, \tau_{12})$ curves are contained between these two limits,
the minimum is present on all of the curves. 

\ifprp
  \begin{figure}[t]
  \epsfxsize=3.5in
  \epsffile{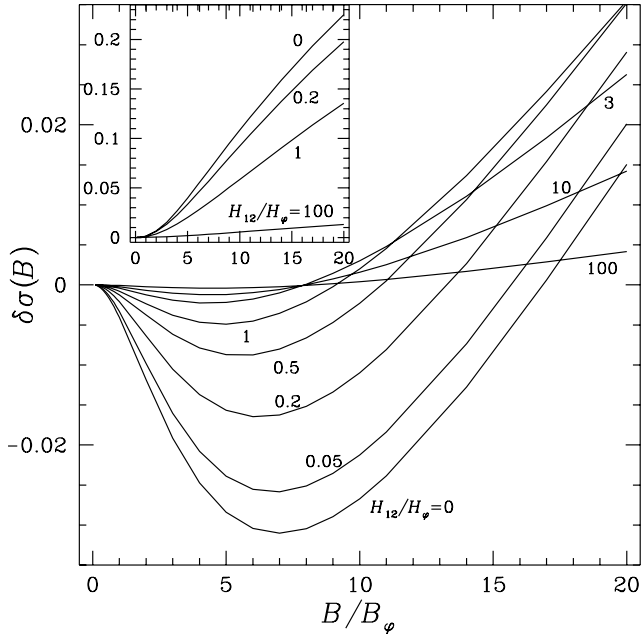}
  \caption{Dimensionless magnetoconductivity $\delta\sigma$ for $N=\infty$,
  $H_{\rm SO}/H_\varphi=2$, and $H_{12}/H_\varphi$ from 0 to 100. 
  The inset shows magnetoconductivity for $H_{\rm SO}/H_\varphi=0.5$.}
  \label{Fig:NInf}
  \end{figure}
\fi

The situation at $N=\infty$ is slightly different, since from the analysis of 
the limiting cases and Eq.~(\ref{Sigma0Inf}) it follows that the conductivity
``per one well'' $\delta\sigma$ tends to 0 at small $\tau_{12}$ (the 
conductivity of the entire system is, of course, infinite for any $\tau_{12}
> 0$). Therefore, with decreasing $\tau_{12}$ the magnetoconductivity
also decreases, as demonstrated in Fig.~\ref{Fig:NInf}. However, the main
qualitative features of the magnetoconductivity dependence on tunneling 
remain the same: if the minimum of $\delta\sigma$ exists with no tunneling,
it also exists at any tunneling rate. The insert shows that the opposite is 
also true: if the dependence $\delta\sigma(B)$ was monotonic for $\tau_{12} =
\infty$, it remains so for any finite $\tau_{12}$.

In conclusion, we have analyzed the effect of weak tunneling between quantum 
wells in a multi-quantum-well structure on the weak localization and the 
magnetoconductivity caused by it. We have found an analytical solution of 
the problem for the case when spin relaxation can be described by a single 
spin relaxation time, neglecting the effects of the linear terms in 
spin relaxation Hamiltonian. The tunneling across the MQW structure was found
to decrease the overall magnitude of the weak localization correction.
However, it does not change the qualitative character of the 
magnetoconductivity, namely, presence (or absence) of regions of positive 
and negative magnetoconductivity. Our results show that the
weak localization correction to the conductivity
in multi-quantum-well structures is sensitive to the
tunneling time $\tau_{12}$ and it should be possible to determine this
time experimentally from the magnetoconductivity measurements.

G. E. P. acknowledges
support by RFFI Grant 96-02-17849 and by the Volkswagen Foundation.

\ifprp
\else
  \eject
  \begin{figure}
  \caption{Cooperon for multi-quantum-well structure, with indices $i$, $j$, $k$,
  and $l$ denoting quantum well numbers.}
  \label{Fig:Cooperon}
  \end{figure}

  \begin{figure*}
  \caption{The system of Dyson equations for
  ${\rm  \kern.24em \vrule width.05em
  height1.4ex depth-.05ex  \kern-.26em C}_{11}$,
  ${\rm  \kern.24em \vrule width.05em
  height1.4ex depth-.05ex  \kern-.26em C}_{21}$, etc.
  in diagrammatic representation. Only first two equations are shown.}
  \label{Fig:Dyson}
  \end{figure*}

  \begin{figure}
  \caption{Function $F^*_N(x)$, which determines Cooperon components entering the 
  expressions for conductivity {\bf (}see Eq.~(\ref{SLF}){\bf )} for $N=2$, 3, 
  4, 6, and $N=\infty$ (solid curves, from highest to lowest). All functions 
  shown by solid curves are obtained without using periodical boundary conditions.
  Using these boundary conditions for $N=4$ results in $F_4(x)$
  shown by short-dash line. Note 
  that the periodical boundary conditions become much more accurate if the $x$
  axis is scaled by $(N-1)/N$ (for the case of $N=4$, shown by long-dashes line,
  the abscissae axis was scaled by $3/4$.}
  \label{Fig:F}
  \end{figure}

  \begin{figure}
  \caption{Dimensionless magneconductivity $\delta\sigma(B)$ for 
  $H_{\rm SO}/H_\varphi = 2$ and $H_{12}/H_\varphi = 4$ and different number 
  of wells $N=\infty$, 6, 4, 3, and $N=2$ (solid lines, in order from higher 
  to lower near the minimum). The solid lines were calculated without using 
  the periodical boundary conditions. The magnetoconductivity calculated with 
  these conditions is shown for $N=4$ by a dashed line.
  The inset shows the effect of spin relaxation on magnetoconductivity for $N=2$
  and $H_{12}/H_\varphi = 4$.}
  \label{Fig:ManyN}
  \end{figure}

  \begin{figure}
  \caption{Dimensionless magnetoconductivity $\delta\sigma$ for $N=3$,
  $H_{\rm SO}/H_\varphi=2$, and $H_{12}/H_\varphi$ from 0 to 100. The 
  squares show $3\cdot \delta\sigma(B, H_{12}/H_\varphi=100)$ (which is
  very close to the limit at $H_{12} = \infty$), to illustrate
  the connection Eq.~(\ref{Sigma0Inf}) between magnetoconductivity at
  $H_{12}=0$ and $H_{12} = \infty$.}
  \label{Fig:N3}
  \end{figure}

  \begin{figure}
  \caption{Dimensionless magnetoconductivity $\delta\sigma$ for $N=\infty$,
  $H_{\rm SO}/H_\varphi=2$, and $H_{12}/H_\varphi$ from 0 to 100. 
  The inset shows magnetoconductivity for $H_{\rm SO}/H_\varphi=0.5$.}
  \label{Fig:NInf}
  \end{figure}

\fi


\begin{references}
\bibitem{LinearTerms} 
It was shown
that for $\rm A_3B_5$ quantum wells, where
the spin relaxation occurs primarily via spin precession due to
spin-orbit terms in the crystal Hamiltonian, the effects of spin
relaxation on the weak localization, in general, cannot be described by
a single relaxation time (and if they can, the time measured from weak
localization will almost certainly be very different from the results of
other, more direct measurements, such as optical orientation
experiments). The description which uses single spin relaxation time is
valid for metal films\cite{Altshuler80}. For $\rm A_3B_5$ quantum wells
it can only be applied in special cases; see
Refs.~\onlinecite{Iordanskii94,Pikus95,Knap96,Pikus96,Hassenkam97}.
\end{references}
\end{document}